\documentclass[letterpaper]{article}
\usepackage{aaai}
\usepackage{times}
\usepackage{helvet}
\usepackage{courier}
\usepackage{graphicx}
\usepackage{amsfonts}
\usepackage{amssymb}
\newtheorem{theorem}{Theorem}
\nocopyright
\begin{document}

\title{Markovian Entanglement Networks}
\author{Pierfrancesco La Mura
\and Lukasz Swiatczak\vspace{1mm}\\\hspace{12mm}plamura@hhl.de \hspace{15mm}lukasz.swiatczak@hhl.de\vspace{3mm}\\HHL-Leipzig Graduate School of Management\\Jahnallee 59, 04109 Leipzig, Germany \vspace{1mm}\\(February 7, 2007)}
\maketitle
\begin{abstract}

Graphical models of probabilistic dependencies have been extensively
investigated in the context of classical uncertainty. However, in some domains
(most notably, in computational physics and quantum computing) the nature of
the relevant uncertainty is non-classical, and the laws of classical
probability theory are superseded by those of quantum mechanics. In this paper
we introduce Markovian Entanglement Networks (MEN), a novel class of graphical
representations of quantum-mechanical dependencies in the context of such
non-classical systems. MEN are the quantum-mechanical analogue of Markovian
Networks, a family of undirected graphical representations which, in the
classical domain, exploit a notion of conditional independence among subsystems.

After defining a notion of conditional independence appropriate to our domain
(\textit{conditional separability}), we prove that the conditional
separabilities induced by a quantum-mechanical wave function are effectively
reflected in the graphical structure of MEN. Specifically, we show that for
any wave function there exists a MEN which is a \textit{perfect map} of its
conditional separabilities. Next, we show how the graphical structure of MEN
can be used to effectively classify the pure states of three-qubit systems. We
also demonstrate that, in large systems, exploiting conditional independencies
may dramatically reduce the computational burden of various inference tasks.
In principle, the graph-theoretic representation of conditional independencies
afforded by MEN may not only facilitate the classical simulation of quantum
systems, but also provide a guide to the efficient design and complexity
analysis of quantum algorithms and circuits.
\end{abstract}

\section{Introduction}

Probabilistic inference in large systems is a topic of great interest in
computer science. In contexts where the underlying uncertainty is of classical
nature, the graphical representation of probabilistic dependencies and
independencies has proved to be a valuable tool in order to reduce the
computational burden of various inference tasks (Pearl 1988). Two main classes
of graphical representations are of special interest in the classical domain.
Bayesian Networks (BN) are based on directed, acyclic graphs which capture
causal dependencies among variables of interest. By contrast, Markovian
Networks\footnote{The word \textit{Markovian} in graphical representations
does not relate directly to Markov processes (a family of stochastic processes
in which the probability distribution of future states for a given history
only depends upon the current state, \textit{i.e.}, future states are
conditionally independent of past states given the present state) but rather
to the Markovian property of conditional independence of a random variable
from all other variables given its neighbors.} (MN) are non-causal
representations based on undirected graphs, which encode mutual symmetric
dependencies among the variables of interest. Whereas the BN formalism is
especially convenient at the modeling stage, as causal dependencies among
variables of interest are more easily identified, the structure of MN is
particularly advantageous from a computational viewpoint. In particular,
several algorithms for probabilistic inference in BN require as a first step
the transformation of the original BN into a corresponding MN (\textit{e.g.,}
the junction tree algorithm).

Recently, there has been some interest in the graphical representation of
non-classical uncertain domains, most notably in the context of computational
physics and quantum computing. (Tucci 1997) introduced Quantum Bayesian
Networks (QBN), which provide a structured representation of
quantum-mechanical wave functions along the lines of classical BN.

Missing a quantum-mechanical generalization of MN, the algorithmic toolbox of
QBN is significantly more limited than that of BN. Also motivated by such
limitations, we introduce a novel class of undirected networks, Markovian
Entanglement Networks (MEN), which extends the formalism and computational
gains of MN to quantum-mechanical systems.

We also introduce a novel notion of conditional independence appropriate to
the context of entangled systems (\textit{conditional separability)}, which
generalizes the classical notion of conditional probabilistic independence,
and provides a rigorous semantics for our graphical framework.

While in applications which only involve classical uncertainty the
computational performance of MEN is equivalent to that of MN, many of the
computational benefits of the modular structure of MN carry over to MEN even
in applications characterized by non-classical uncertainty. In particular, the
graphical structure of MEN can in principle be exploited in the context of
classical simulations of quantum-mechanical systems (\textit{e.g.,} protein
folding), and as a guide to the efficient design and complexity analysis of
quantum algorithms and circuits.

The paper is organized as follows. In the second section we introduce
conditional separability, and relate it with both separability and the
classical notion of conditional probabilistic independence. In the third
section we formally introduce MEN, and discuss some of their structural
properties. In particular, we show that the conditional separabilities induced
by any wave function are precisely characterized by the graphical structure of
an appropriately selected MEN. In the fourth section we address the issue of
probabilistic inference, showing how marginal and conditional probabilities
can be recovered from the structural elements of MEN, and
discussing the computational performance of MEN\ in the context of some
typical inference tasks. The fifth section provides a classification of
3-qubit states based on their network topologies, while the last section
concludes. 
\section{Setting, \textit{a-}Independence, Separability, Conditional Separability}

\vspace{2mm}Let us first introduce the notion of a \emph{quantum bit}
(\emph{qubit}), the fundamental unit of quantum information, which can be
thought of as a quantum mechanical analogue of a classical data bit. A
\emph{(pure) state} of a quantum bit is represented by a vector in a two-dimensional
complex Hilbert space. The two computational basis states are conventionally
written as $\left|  0\right\rangle $ and $\left|  1\right\rangle ,$ and form
an orthonormal basis\footnote{An orthonormal basis of a vector space $V$
equipped with inner product $\left\langle \cdot,\cdot\right\rangle $ is a
subset $\left\{  v_{1},...,v_{k}\right\}  $ of $V$ which spans the whole
space, and is such that $\left(  \forall_{{}}i,j\right)  \left[  \left\langle
v_{i},v_{j}\right\rangle =\delta_{ij}\right]  $, where $\delta_{ij}$ is the
Kronecker Delta.} for the Hilbert space. A qubit state is a superposition of
those basis states:

\begin{center}%
\[
\left|  \psi\right\rangle =a(\left|  0\right\rangle )\left|  0\right\rangle
+a(\left|  1\right\rangle )\left|  1\right\rangle ,
\]
\end{center}

\vspace{2mm}where $a(\left|  0\right\rangle )$ and $a(\left|  1\right\rangle
)$ are complex numbers denoting probability amplitudes in the directions given
by $\left|  0\right\rangle $ and $\left|  1\right\rangle $, respectively.
These amplitudes are normalized so that their Euclidean norm is unitary,
\textit{i.e.,} $\left|  a(\left|  0\right\rangle )\right|  ^{2}+\left|
a(\left|  1\right\rangle )\right|  ^{2}=1.$ The reason is that the
probabilities that the qubit will be observed in states $\left|
0\right\rangle $ and $\left|  1\right\rangle ,$ respectively, are represented
by the square moduli $\left|  a(\left|  0\right\rangle )\right|  ^{2}$ and
$\left|  a(\left|  1\right\rangle )\right|  ^{2}$, so with normalized
amplitudes the total probability that the qubit is observed to be in either
state $\left|  0\right\rangle $ or $\left|  1\right\rangle $ is $1$. For
notational convenience, in the remainder of this paper we shall give up the
bra-ket notation $\left|  \cdot\right\rangle $ and always write $x$ in place
of $\left|  x\right\rangle $. \vspace{2mm}\newline A pure state
(\textit{wave function}) of an \emph{n-qubit system} takes the following form:%

\[
\psi_{N}=\sum_{x\in X}a(x)x,
\]
\vspace{2mm}where: \newline 

\begin{itemize}
\item $N=\left\{  1,...,n\right\}  $ is a set of $n$ qubits,

\item $x=x_{1}\otimes...\otimes\ x_{n}$ is a vector in an orthonormal basis
$X$ of the $n$-qubit system, while each $x_{i}\in\left\{  0_{i},1_{i}\right\}
$ is an orthonormal basis of the $i-th$ qubit, $i=1,...,n$, and \footnote{The
notation $v\otimes u$ stands for the tensor product of two vectors $v$ and $u$.}

\item $a(x)$ is a complex number denoting the probability amplitude of the
$n$-qubit system in the direction given by $x$, normalized so that $\sum_{x\in X}\left|
a(x)\right|  ^{2}=1$.
\end{itemize}
In this paper we only deal with pure states, while deferring the case of mixed states\footnote{A mixed state is a classical
probability distribution over pure states. Mixed states are conventionally
represented by a density matrix $\rho$ (\textit{i.e.,} a self-adjoint,
positive-semidefinite matrix of trace one) with $Tr(\rho^{2})<1$.} to future work.

The state of a composite system is said to be \textit{separable} (or \textit{not
entangled}) if it can be written as tensor product of states of
the component systems (Nielsen and Chuang 2000). Formally,
\[
\psi_{N}=\psi_{M,N-M}=\psi_{M}\otimes\psi_{N-M}.
\]
Intuitively, the subsystems do not interact with each other, so they can be considered separately. This
implies probabilistic independence between subsets of random variables across
such non-interacting subsystems. 
\vspace{2mm}\\ Let $\left\{  X_{i}\right\}  _{i\in N}$ be a finite, ordered set
of Boolean random variables representing the possible realizations of the $n$
qubits, and let $x_{N}{}^{0}=(x_{1}^{0},...,x_{n}^{0})$, where $x_{i}{}^{0}$
is a given realization of the $i-th$ random variable, be an arbitrary
reference point.\footnote{We always use capital letters to denote random
variables, and small letters to denote their realizations.}. A joint
realization $x_{N}=(x_{1},...,x_{n})$ corresponds to a \textit{basis state} of
the composite $n-$qubit system. For any $M\subset N$ we denote by$\ \overline
{M}$ the set $N-M,$ and by $X_{M}$ the set $\left\{  X_{i}\right\}  _{i\in M}.$

\vspace{1mm}Finally, we define a notion of $a-$\textit{independence} which, as we shall
see, is equivalent to separability. \vspace{1mm}\newline Two sets of qubits, $M$ and $\overline{M}$, are said to be $a$-\textit{independent} if the
following condition is satisfied for all $x_{M}$ and $x_{\overline{M}}$:

\begin{center}%
\[
a(x_{M},x_{\overline{M}})a(x_{M}^{0},x_{\overline{M}}^{0})=a(x_{M}%
^{0},x_{\overline{M}})a(x_{M},x_{\overline{M}}^{0}).\vspace{3mm}%
\]
\end{center}

\begin{theorem}
Two subsystems $M$ and $\overline{M}$ are $a$-independent if and only if they
are separable. Formally, \vspace{3mm}$\newline \forall(x_{N})\left[
a(x_{M},x_{\overline{M}})a(x_{M}^{0},x_{\overline{M}}^{0})=a(x_{M}%
^{0},x_{\overline{M}})a(x_{M},x_{\overline{M}}^{0})\right]  $
\begin{flushright}
$\Leftrightarrow\psi_{N}=\psi_{M}\otimes\psi_{\overline{M}}$.
\end{flushright}
\end{theorem}

\vspace{2mm}\textbf{Proof.} We prove \textit{Theorem 1} through a two-step
procedure. We first show that separability of the two subsystems implies their
\textit{a}-independence. Next we show the converse, namely that \textit{a}%
-independence implies separability. The two steps are sufficient to conclude
the proof. \newline \newline \textit{Step 1}. \textit{separability} $\Rightarrow$
\textit{a-independence} \vspace{2mm}\newline If the two subsystems, $M$
and $\overline{M}$, are separable then $\psi_{N}=\psi
_{M}\otimes\psi_{\overline{M}}$. Let us write each of these wave functions in
vector form:

\begin{center}%
\[
\psi_{F}= \left(
\begin{array}
[c]{c}%
a(x_{F}^{1})\\
\vdots\\
a(x_{F}^{2^{\left|  F\right|  }})
\end{array}
\right)
\]
\end{center}

where $\left|  F\right|  $ is the cardinality of set $F$, and $2^{\left|
F\right|  }$ is the number of basis states of the $|F|-$qubit system.
\newline Next, we rewrite the separability condition in the following form

\[
\left(
\begin{array}
[c]{c}%
a(x_{N}^{1})\\
\vdots\\
a(x_{N}^{2^{\left|  N\right|  }})
\end{array}
\right)  = \left(
\begin{array}
[c]{c}%
a(x_{M}^{1})\\
\vdots\\
a(x_{M}^{2^{\left|  M\right|  }})
\end{array}
\right)  \otimes\left(
\begin{array}
[c]{c}%
a(x_{\overline{M}}^{1})\\
\vdots\\
a(x_{\overline{M}}^{2^{\left|  \overline{M}\right|  }})
\end{array}
\right)
\]
or, equivalently,
\[
\left(
\begin{array}
[c]{c}%
a(x_{M}^{1},x_{\overline{M}}^{1})\\
\vdots\\
a(x_{M}^{1}, x_{\overline{M}}^{2^{\left|  \overline{M}\right|  }})\\
\vdots\\
a(x_{M}^{2^{\left|  M\right|  }},x_{\overline{M}}^{2^{\left|  \overline
{M}\right|  }})
\end{array}
\right)  = \left(
\begin{array}
[c]{c}%
a(x_{M}^{1})a(x_{\overline{M}}^{1})\\
\vdots\\
a(x_{M}^{1})a(x_{\overline{M}}^{2^{\left|  \overline{M}\right|  }})\\
\vdots\\
a(x_{M}^{2^{\left|  M\right|  }})a(x_{\overline{M}}^{2^{\left|  \overline
{M}\right|  }})
\end{array}
\right)
.\]
\vspace{2mm}It is straightforward to verify that the following condition must
then hold:
\[
a(x_{M}^{i},x_{\overline{M}}^{j})a(x_{M}^{i^{\prime}},x_{\overline{M}%
}^{j^{\prime}})=a(x_{M}^{i^{\prime}},x_{\overline{M}}^{j})a(x_{M}%
^{i},x_{\overline{M}}^{j^{\prime}})
\]
\newline for all $x_{M}^{i},x_{M}^{i^{\prime}},x_{\overline{M}}^{j}$ and
$x_{\overline{M}}^{j\prime}$. In particular, the condition must hold true when
$x_{M}^{i^{\prime}}=x_{M}^{0}$ and $x_{\overline{M}}^{j^{\prime}}%
=x_{\overline{M}}^{0}.$ It follows that separability implies \textit{a}%
-independence. 
\newline \newline \textit{Step 2}. \textit{a-independence}$\Rightarrow$ \textit{separability} 
\\\\By $a-$independence of systems $M$ and $\overline{M}$, we know that the following condition holds true for all $x_{M}$ and $x_{\overline{M}}$:
\[
a(x_{M}^{i},x_{\overline{M}}^{0})a(x_{M}^{0}%
,x_{\overline{M}}^{j})=a(x_{M}^{i},x_{\overline{M}}^{j})a(x_{M}^{0},x_{\overline{M}}^{0}).
\]
This can then be rewritten as
\[
a(x_{M}^{i},x_{\overline{M}}^{j})=\frac{a(x_{M}^{i},x_{\overline{M}}^{0})a(x_{M}^{0}%
,x_{\overline{M}}^{j})}{a(x_{N}^{0})},
\]
where $a(x_{N}^{0})$ stands for the amplitude of the arbitrary reference point
$x_{N}^{0}$, which without loss of generality we assume to be non-zero.
\vspace{1mm}\\Thus, the state of the $n-$qubit system takes the form
\begin{flushleft}
$\psi_{N}=
\left(
\begin{array}
[c]{c}%
a(x_{M}^{1},x_{\overline{M}}^{1})\\
\vdots\\
a(x_{M}^{1},x_{\overline{M}}^{2^{\left|  \overline{M}\right|  }})\\
\vdots\\
a(x_{M}^{2^{\left|  M\right|  }},x_{\overline{M}}^{2^{\left|  \overline
{M}\right|  }})
\end{array}
\right)$
\end{flushleft}
\begin{flushleft}
\hspace{20mm}=$\frac{1}{a(x_{N}^{0})}
\left(
\begin{array}
[c]{c}
a(x_{M}^{1},x_{\overline{M}}^{0})a(x_{M}^{0}%
,x_{\overline{M}}^{1})\\
\vdots\\
a(x_{M}^{1},x_{\overline{M}}^{0})a(x_{M}^{0}%
,x_{\overline{M}}^{2^{\left|\overline{M}\right|}})\\
\vdots\\
a(x_{M}^{2^{|M|}},x_{\overline{M}}^{0})a(x_{M}^{0}%
,x_{\overline{M}}^{2^{\left|\overline{M}\right|}})\\
\end{array}
\right)$
\end{flushleft}
which factorizes in the following way
		\[
		\psi_{N}=\frac{1}{a(x_{N}^{0})}
			\left(\begin{array}{c}
			a(x_{M}^{1},x_{\overline{M}}^{0}) \\ \vdots\\ a(x_{M}^{2^{\left|M\right|}},x_{\overline{M}}^{0})
			\end{array} \right) \otimes
				\left( \begin{array}{c}
					a(x_{M}^{0},x_{\overline{M}}^{1}) \\ \vdots\\ a(x_{M}^{0}, x_{\overline{M}}^{2^{\left|\overline{M}\right|}})
				\end{array} \right). 
		\]
Let us denote the vectors in the above tensor product as $\alpha$ and $\beta$, and rewrite the last equation as follows:
\[
\psi_{N}=\frac{1}{a(x_{N}^{0})}\alpha\otimes\beta.
\]
Let $c_{\alpha}, c_{\beta}$ be complex numbers such that $c_{\alpha}c_{\beta}=\frac{1}{a(x_{N}^{0})}$, and $\left|c_{\alpha}\right|\left\|\alpha\right\|=\left|c_{\beta}\right|\left\|\beta\right\|$. 
It follows that
\[
\psi_{N}=\alpha'\otimes\beta',
\]
where $\alpha'=c_{\alpha}\alpha$, $\beta'=c_{\beta}\beta$.
Taking the norm on both sides, and using the fact that $\psi_{N}$ has unit norm, we find that 
\[
\left\|\alpha'\right\| \left\|\beta'\right\|=1.
\] 
This implies $\left\|\alpha'\right\|=\left\|\beta'\right\|=1$ (\textit{i.e.,}  $\alpha'$, and $\beta'$ are pure states of the component systems).$\blacksquare$ 
\vspace{2mm} \newline The above characterization of
separability via $a-$independence makes it possible to introduce in a natural
way a notion of \textit{conditional separability}, defined as follows. Let
$\left\{  A,B,C\right\}  $ be a partition of $N$ into non-empty subsets. We
say that $A$ and $B$ are \textit{conditionally separable} given $C$ if
\vspace{2mm}\newline $\forall(x_{N})[a(x_{A},x_{B},x_{C})a(x_{A}^{0},x_{B}%
^{0},x_{C})$ \begin{flushright}
$=a(x_{A}^{0},x_{B},x_{C})a(x_{A},x_{B}^{0},x_{C})].$
\end{flushright}
Whenever two subsystems are not conditionally separable, we say that they are \textit{conditionally entangled}.

\begin{theorem}
Conditional separability implies conditional probabilistic independence. \newline 
\end{theorem}

\textbf{Proof.} Observe that conditional probabilistic independence of $X_{A}$
and $X_{B}$ given $X_{C}$ can be expressed as%
\[
\forall(x_{N})\left[  p(x_{A}|x_{B},x_{C})=p(x_{A}|x_{B}^{0},x_{C})\right]  ,
\]
where%
\[
p(x_{A}|x_{B},x_{C})=\frac{p(x_{A},x_{B},x_{C})}{\sum_{x_{A}}p(x_{A}%
,x_{B},x_{C})}.\vspace{1mm}%
\]
In terms of probability amplitudes, this condition takes the following form:%
\[
p(x_{A}|x_{B},x_{C})=\frac{\left|  a(x_{A},x_{B},x_{C})\right|  ^{2}}%
{\sum_{x_{A}}\left|  a(x_{A},x_{B},x_{C})\right|  ^{2}}\vspace{1mm}.
\]
By multiplying the above fraction's numerator and denominator by $\left|
a(x_{A}^{0},x_{B}^{0},x_{C})\right|  ^{2}$ and by the fact that $|c_{1}%
||c_{2}|=|c_{1}c_{2}|$, where $c_{1},c_{2}\in\mathbb{C}$, we obtain
\vspace{1mm}
\[
p(x_{A}|x_{B},x_{C})=\frac{|a(x_{A},x_{B},x_{C})a(x_{A}^{0},x_{B}^{0}%
,x_{C})|^{2}}{\sum_{x_{A}}|a(x_{A},x_{B},x_{C})a(x_{A}^{0},x_{B}^{0}%
,x_{C})|^{2}}.\vspace{1mm}%
\]
Next, from conditional separability of $A$ and $B$ given $C$, one obtains
that\vspace{1mm}
\[
p(x_{A}|x_{B},x_{C})=\frac{|a(x_{A}^{0},x_{B},x_{C})a(x_{A},x_{B}^{0}%
,x_{C})|^{2}}{\sum_{x_{A}}|a(x_{A}^{0},x_{B},x_{C})a(x_{A},x_{B}^{0}%
,x_{C})|^{2}}\vspace{1mm}%
\]%
\[
\hspace{18mm}=\frac{|a(x_{A}^{0},x_{B},x_{C})|^{2}|a(x_{A},x_{B}^{0}%
,x_{C})|^{2}}{|a(x_{A}^{0},x_{B},x_{C})|^{2}\sum_{x_{A}}|a(x_{A},x_{B}%
^{0},x_{C})|^{2}}\vspace{1mm}%
\]%
\[
\hspace{38mm}=\frac{|a(x_{A},x_{B}^{0},x_{C})|^{2}}{\sum_{x_{A}}|a(x_{A}%
,x_{B}^{0},x_{C})|^{2}}.\vspace{1mm}%
\]
Observing that the latter term is equal to $p(x_{A}|x_{B}^{0},x_{C})$
concludes the proof. $\blacksquare$ 
\vspace{2mm}\newline Finally, we introduce
a potential function that we shall utilize in defining Markovian Entanglement
Networks. Following the approach of (La Mura and Shoham 1999), we define the
\emph{amplitude potential function} (\textit{$q-$function}) as follows:

\[
q(x_{M}|x_{\overline{M}})=\frac{a(x_{M},x_{\overline{M}})}{a(x_{M}%
^{0},x_{\overline{M}})}.
\]

\vspace{2mm}The \textit{$q$-}function\textit{,} whenever it is well defined,
can be interpreted in terms of \textit{ceteris paribus} comparisons: it
describes how the probability amplitude changes when the realizations of the qubits in $M$ are
shifted away from the reference point, while those in
$\overline{M}$ are held unchanged at $x_{\overline{M}}.$ \vspace{2mm}%
\newline Note that, whenever $q$ is well defined, the condition $\forall
(x_{N})[q(x_{A}|x_{B},x_{C})=q(x_{A}|x_{B}^{0},x_{C})]$ exactly corresponds to
conditional separability of $A$ and $B$ given $C$, in which case we shall
simply write $q(x_{A}|x_{C})$ in place of $q(x_{A}|x_{B},x_{C})$.

\section{Markovian Entanglement Networks: a formal definition}

\vspace{2mm}We define a Markovian Entanglement Network as an undirected graph
$G(V,E)$ with nodes $i\in V$ representing quantum-mechanical
subsystems, and edges $\left\{  i,j\right\}  \in E$ representing conditional entanglement between
subsystems $i$ and $j$. For simplicity of exposition, we proceed under the
assumption that each node represents a single quantum bit. Finally, each node is associated with a non-zero function
$q(x_{i}|x_{U(i)})$ (as defined in the previous section), where $U(i)$ denotes
the set of nodes directly connected to $i$ via
entanglement edges (or\emph{ neighbors}). Figure 1 depicts a simple MEN.

\begin{figure}[pth]
\begin{center}
\includegraphics[width=2.5in]{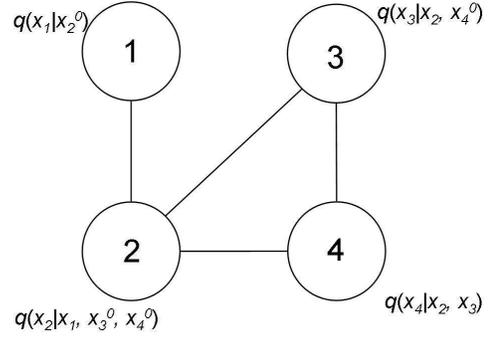}
\end{center}
\caption{A simple MEN. Nodes represent qubits, and edges the fact that two
qubits are conditionally entangled.}%
\end{figure}

If the $q-$functions are specified directly, then any arbitrary assignment of
non-zero complex-valued functions $q(x_{i}|x_{U_{-}(i)},x_{U_{+}(i)}^{0})$ for
all $i$ (where $U_{+}(i)$ denotes the set of all qubits in $U(i)$ whose index
is greater than $i$, and $U_{-}(i)=U(i)-U_{+}(i))$ uniquely identifies a
corresponding wave function, up to the phase of the reference point. In fact,
once all the $q-$functions are specified, the relative amplitudes
$a(x_{N})/a(x_{N}^{0})$ are determined as follows: \vspace{2mm}\newline
$\frac{a(x_{N})}{a(x_{N}^{0})}=\frac{a(x_{1},x_{R}^{0})}{a(x_{N}^{0})}%
\frac{a(x_{N})}{a(x_{1},x_{R}^{0})}\vspace{2mm}\newline =q(x_{1}|x_{U(i)}%
^{0})\frac{a(x_{1},x_{2},x_{R^{\prime}}^{0})}{a(x_{1},x_{2}^{0},x_{R^{\prime}%
}^{0})}\frac{a(x_{N})}{a(x_{1},x_{2},x_{R^{\prime}}^{0})}$\vspace
{2mm}\newline $=q(x_{1}|x_{U_{+}(i)}^{0})q(x_{2}|x_{1},x_{U_{+}(i)}^{0}%
)\frac{a(x_{N})}{a(x_{1},x_{2},x_{R^{\prime}}^{0})}$ \vspace{2mm}%
\newline $=...=\prod_{i}q(x_{i}|x_{U_{-}(i)},x_{U_{+}(i)}^{0}),$ \vspace
{2mm}\newline where $x_{R},x_{R^{\prime}}$ denote the realizations of the
remaining qubits in the appropriate context. \vspace{2mm}\newline Since the
state of the composite $n$\textit{-qubit} system is pure, we obtain the
following condition: \vspace{1mm}

\begin{center}
$\sum_{x_{N}}\left|  \prod_{i}q(x_{i}|x_{U_{-}(i)},x_{U_{+}(i)}^{0}%
)a(x_{N}^{0})\right|  ^{2}=1.$
\end{center}

Performing some elementary algebra, we conclude that
\[
\left|  a(x_{N}^{0})\right|  =\frac{1}{\sqrt{\sum_{x_{N}}\left|  \prod
_{i}q(x_{i}|x_{U_{-}(i)},x_{U_{+}(i)}^{0})\right|  ^{2}}}.
\]

From the last condition follows that only the the modulus, but not the phase,
of the reference amplitude $a(x_{N}^{0})$ is identified, which means that the
corresponding wave function is only determined up to the phase of the
reference point. Conversely, any wave function in a given computational basis
can be represented by an appropriately selected MEN, as long as all the
amplitudes are non-zero. \newline We remark that, if $q(x_{i}|x_{N-\left\{
i\right\}  })$ only depends on $x_{U(i)},$ then fixing $x_{U(i)}$ completely
specifies the behavior of the wave function along the $i-th$ coordinate (up to
the phase of the reference point), and that such behavior does not depend on
the specific values taken by the remaining variables. \hspace{2mm}\newline It
turns out that node separation with respect to MEN characterizes all the
implied conditional separabilities. More precisely, for any wave function
there exists an undirected graph $G$ such that, for any partition of $N$ into
three non-empty sets of quantum bits $A$, $B$, and $C$, $A$ is conditionally
separable from $B$ given $C$ if, and only if, $C$ separates $A$ from $B$ in
graph $G$, \textit{i.e.}, every path from a node in $A$ to a node in $B$
passes through some node in $C$. In the terminology of (Pearl and Paz 1987),
such a graph is said to be a \emph{perfect map} of the independence structure.

\begin{theorem}
The set of conditional separabilities generated by any wave function has a
perfect map. \newline 
\end{theorem}

\textbf{Proof.} We appeal to a necessary and sufficient condition in (Pearl
and Paz 1987). Specifically, we check whether conditional separability
satisfies the following five properties: symmetry, decomposition,
intersection, strong union and transitivity. \newline Let $A$, $B$, $C$, $D$,
$R$, $R^{\prime}$, $R^{\prime\prime}$ be subsets of qubits, where
$R,R^{\prime},R^{\prime\prime}$ denote the subset of remaining qubits in the
appropriate context. \newline For the purpose of this proof, let us say that
$A$ is conditionally independent of $B$ given $C$, and write $I(A,B|C)$ if and
only if
\[
\hspace{-20mm}(\forall x_{N})[a(x_{A},x_{B},x_{C},x_{R})a(x_{A}^{0},x_{B}%
^{0},x_{C},x_{R})
\]%
\[
\hspace{28mm}=a(x_{A}^{0},x_{B},x_{C},x_{R})a(x_{A},x_{B}^{0},x_{C},x_{R})].
\]
Observe that this formulation reduces to conditional separability if $A, B, C$ are a partition of $N$.
\\Then the following properties hold. \newline \newline \textbf{Symmetry:}
$I(A,B|C)\Rightarrow I(B,A|C).$ \vspace{2mm}\newline This follows trivially
from the definition of conditional independence. Specifically,
$I(A,B|C)\Leftrightarrow I(B,A|C).$
%
%
%
\newline \newline \textbf{Decomposition:} $I(A,B\cup D|C)\Rightarrow
I(A,B|C)\wedge I(A,D|C).$ \vspace{2mm}\newline This is equivalent to
\[
\hspace{-13mm}[a(x_{A},x_{B},x_{C},x_{D},x_{R})a(x_{A}^{0},x_{B}^{0}%
,x_{C},x_{D}^{0},x_{R})=
\]%
\[
\hspace{12mm}a(x_{A}^{0},x_{B},x_{C},x_{D},x_{R})a(x_{A},x_{B}^{0},x_{C}%
,x_{D}^{0},x_{R})]\Rightarrow\vspace{2mm}%
\]
%
%
%
\[
\hspace{-20mm}[a(x_{A},x_{B},x_{C},x_{R^{\prime}})a(x_{A}^{0},x_{B}^{0}%
,x_{C},x_{R^{\prime}})=
\]%
\[
\hspace{20mm}a(x_{A}^{0},x_{B},x_{C},x_{R^{\prime}})a(x_{A},x_{B}^{0}%
,x_{C},x_{R^{\prime}})]\hspace{2mm}\wedge
\]%
\[
\hspace{-20mm}[a(x_{A},x_{C},x_{D},x_{R^{\prime\prime}})a(x_{A}^{0}%
,x_{C},x_{D}^{0},x_{R^{\prime\prime}})=
\]%
\[
\hspace{25mm}a(x_{A}^{0},x_{C},x_{D},x_{R^{\prime\prime}})a(x_{A},x_{C}%
,x_{D}^{0},x_{R^{\prime\prime}})].
\]
\vspace{3mm}This follows trivially, since $R^{\prime}=(D,R)$ and
$R^{\prime\prime}=(B,R)$.
\newline \textbf{Intersection:} \\$I(A,B|C\cup D)\wedge I(A,D|B\cup
C)\Rightarrow I(A,B\cup D|C).$\vspace{2mm} \newline Equivalently,
\[
\hspace{-14mm}[a(x_{A},x_{B},x_{C},x_{D},x_{R})a(x_{A}^{0},x_{B}^{0}%
,x_{C},x_{D},x_{R})=
\]%
\[
\hspace{12mm}a(x_{A}^{0},x_{B},x_{C},x_{D},x_{R})a(x_{A},x_{B}^{0},x_{C}%
,x_{D},x_{R})]\hspace{2mm}\wedge
\]%
\[
\hspace{-14mm}[a(x_{A},x_{B},x_{C},x_{D},x_{R})a(x_{A}^{0},x_{B},x_{C}%
,x_{D}^{0},x_{R})=
\]%
\[
\hspace{10mm}a(x_{A}^{0},x_{B},x_{C},x_{D},x_{R})a(x_{A},x_{B},x_{C},x_{D}%
^{0},x_{R})]\hspace{2mm}\Rightarrow
\]%
\[
\hspace{-14mm}[a(x_{A},x_{B},x_{C},x_{D},x_{R})a(x_{A}^{0},x_{B}^{0}%
,x_{C},x_{D}^{0},x_{R})=
\]%
\[
\hspace{10mm}a(x_{A}^{0},x_{B},x_{C},x_{D},x_{R})a(x_{A},x_{B}^{0},x_{C}%
,x_{D}^{0},x_{R})].
\]
This follows quite easily by algebraic manipulation. \vspace{3mm}%
\newline \textbf{Strong union:} $I(A,B|C)\Rightarrow I(B,A|C\cup D).$
\vspace{1mm} \newline Equivalently,\vspace{1mm}
\[
\hspace{-26mm}[a(x_{A},x_{B},x_{C},x_{R})a(x_{A}^{0},x_{B}^{0},x_{C},x_{R})=
\]%
\[
\hspace{14mm}a(x_{A}^{0},x_{B},x_{C},x_{R})a(x_{A},x_{B}^{0},x_{C}%
,x_{R})]\hspace{2mm}\Rightarrow
\]%
\[
\hspace{-12mm}[a(x_{A},x_{B},x_{C},x_{D},x_{R^{\prime}})a(x_{A}^{0},x_{B}%
^{0},x_{C},x_{D},x_{R^{\prime}})=
\]%
\[
\hspace{10mm}a(x_{A}^{0},x_{B},x_{C},x_{D},x_{R^{\prime}})a(x_{A},x_{B}%
^{0},x_{C},x_{D},x_{R^{\prime}})].
\]
This follows by symmetry, and the fact that $R=(D,R^{\prime})$.
\newline \newline \textbf{Transitivity:} $I(A,B|C)\Rightarrow I(A,V|C)\vee
I(B,V|C)$\vspace{2mm}\newline where $V$ is one qubit. This is equivalent to
\[
\hspace{-26mm}[a(x_{A},x_{B},x_{C},x_{R})a(x_{A}^{0},x_{B}^{0},x_{C},x_{R})=
\]%
\[
\hspace{10mm}a(x_{A}^{0},x_{B},x_{C},x_{R})a(x_{A},x_{B}^{0},x_{C}%
,x_{R})]\hspace{2mm}\Rightarrow
\]%
\[
\hspace{-24mm}[a(x_{A},x_{C},x_{V},x_{R^{\prime}})a(x_{A}^{0},x_{C},x_{V}%
^{0},x_{R^{\prime}})=
\]%
\[
\hspace{11mm}a(x_{A}^{0},x_{C},x_{V},x_{R^{\prime}})a(x_{A},x_{C},x_{V}%
^{0},x_{R^{\prime}})]\hspace{2mm}\vee
\]
%
%
%
\[
\hspace{-22mm}[a(x_{B},x_{C},x_{V},x_{R^{\prime\prime}})a(x_{B}^{0}%
,x_{C},x_{V}^{0},x_{R^{\prime\prime}})=
\]%
\[
\hspace{12mm}a(x_{B}^{0},x_{C},x_{V},x_{R^{\prime\prime}})a(x_{B},x_{C}%
,x_{V}^{0},x_{R^{\prime\prime}})].
\]
\vspace{2mm}\newline which follows by observing that $V$ is either an element
of $A$, or $B$, or $R.$ Thus, we appeal to Pearl and
Paz's result to conclude the proof.$\blacksquare$ \vspace{2mm}\newline The
relevance of this result lies in the fact that one can represent and reason
about conditional separabilities graphically (\textit{i.e.}, one can exploit
the graphical structure of the perfect map to reason about conditional separabilities).

\section{Exact Probabilistic Inference in MEN}

\vspace{3mm}Markovian Entanglement Networks contain information about quantum
probabilistic dependencies, but the information is not explicit. It is encoded
in the potential functions, together with the topological structure of the
network. The basic operation of computing marginal probabilities, and hence
conditional probabilities, can be reduced to manipulations of the wave
potentials. In fact, one readily obtains $p(x_{M})/p(x_{N}^{0})$ (where
$p(x_{M})$ is the marginal probability of obtaining $x_{M}$) by summing up the
squared moduli of the $q$-functions:

\begin{center}
$\frac{p(x_{M})}{p(x_{N}^{0})}=\sum_{x_{N-M}}\left|  \prod_{i}q(x_{i}%
|x_{U_{-}(i)},x_{U_{+}(i)}^{0})\right|  ^{2}.$
\end{center}

\vspace{2mm} The computational performance of MEN in evaluating probabilistic
queries is similar to that of MN. In particular, as in the case of MN, exact
inference in singly-connected networks\footnote{A network is said to be
\textit{singly connected} if, in the associated graph, there exists at most
one path between any two nodes.} can be performed in time linear in the number
of nodes. We provide time-complexity analyses for some basic inference tasks
in the simplest singly-connected networks, chains (\textit{Figure 2}).

\begin{center}
\begin{figure}[pth]
\begin{center}
\includegraphics[
					width=3in]
					{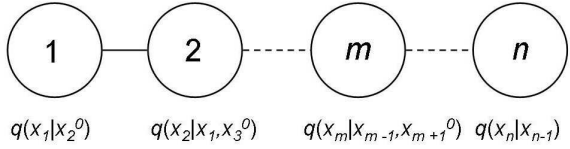}
\end{center}
\caption{A MEN chain.}%
\end{figure}
\end{center}

Specifically, we consider three types of queries: $(i)$ computing marginal
probabilities, $(ii)$ computing conditional probabilities given evidence, and
$(iii)$ finding a maximum likelihood instantiation. \newline \newline $(i)$
Computing marginal probability ratios takes the following form:
\[
\frac{p(x_{M})}{p(x_{N}^{0})}=\sum_{x_{N-M}}\left|  \prod_{i}q(x_{i}%
|x_{i-1},x_{i+1}^{0})\right|  ^{2}.
\]
\newline The right-hand side of the above formula can be rearranged as follows:

\begin{flushleft}
$\sum_{x_{N-M}}\prod_{i=1}^{m}q(x_{i}|x_{i-1},x_{i+1}^{0})q(x_{m+1}%
|x_{m},x_{m+2}^{0}) \cdots$
\end{flushleft}
\begin{flushright}
$\cdots q(x_{n}|x_{n-1})|^{2}$
\end{flushright}
\vspace{2mm} We assume, for simplicity, that $M\subseteq N$ is the set of
qubits with indices from $1$ to $m$, whereas in $N-M$ are all the qubits with
higher indices. Hence, the following holds: \begin{flushleft}
$\sum_{x_{m+1}}\ldots\sum_{x_{n}}\left|\prod_{i=1}^{m}q(x_{i}|x_{i-1}%
,x_{i+1}^{0}) \cdots q(x_{n}|x_{n-1})\right|^{2}=$
\end{flushleft}
\begin{flushleft}
$\prod_{i=1}^{m}|q(x_{i}|x_{i-1},x_{i+1}^{0})|^{2}\sum_{x_{m+1}}%
\left|q(x_{m+1}|x_{m},x_{m+2}^{0})\right|^{2} \cdots$
\end{flushleft}
\begin{flushright}
$\cdots\sum_{x_{n}}\left|q(x_{n}|x_{n-1})\right|^{2}.$
\end{flushright}
\vspace{2mm} The above decomposition makes it possible to compute the marginal
probability of interest performing $6(n-m)+2m-1$ operations. Thus, the time
complexity of this task is linear in the number of qubits $n$. \newline
\newline $(ii)$ Computing conditional probabilities given some evidence
$x_{M}$ takes the form \begin{flushleft}
$p(x_{N-M}|x_{M})=\frac{p(x_{M},x_{N-M})}{p(x_{M})}=\frac{p(x_{M}%
,x_{N-M})/p(x_{N}^{0})}{p(x_{M})/p(x_{N}^{0})}.$
\end{flushleft}
\vspace{2mm} Since the computation of marginal probability ratios has linear
time complexity, this is also the case for the computation of conditional
probabilities. 
\vspace{1mm}\\$(iii)$ Finding a maximum likelihood instantiation takes the form
\[
argmax_{x_{N}}\left|  a\left(  x_{N}\right)  \right|  ^{2}\Leftrightarrow
argmax_{x_{N}}\left|  \prod_{i}q(x_{i}|x_{i-1},x_{i+1}^{0})\right|  ^{2}%
\]
\begin{flushleft}
$=argmax_{x_{1}}\ldots argmax_{x_{n}}\left|q(x_{1}|x_{2}^{0})\cdots
q(x_{n}|x_{n-1})\right|^{2}$
\end{flushleft}
\begin{flushleft}
$= argmax_{x_{1}}\left|q(x_{1}|x_{2}^{0})\right|^{2}\cdots argmax_{x_{n}}%
\left|q(x_{n}|x_{n-1})\right|^{2}$
\end{flushleft}
Thanks to the exploitation of conditional separabilities, the problem enjoys a
decomposition which, once again, allows to find a solution in time linear in
the number of qubits. Specifically, a global maximum can be found by solving a
sequence of $n$ simpler maximization problems.

\section{MEN for 3-Qubit Systems}

From Theorem 3, we know that all conditional separabilities induced by any
given wave function can be encoded in the graphical layer of an appropriately
chosen MEN. In the case of a two-qubit system, the dependencies generated are
trivial: either the two qubits are entangled, or they are separable. Hence, for
two-qubit systems there are only two classes of MEN topologies\textit{:} one
with connected, and one with unconnected nodes. Since in two-qubit systems
conditional separability plays no role, in this section we investigate
conditional separabilities in the context of the simplest interesting class:
three-qubit systems. First, following (Duer, Vidal, and Cirac 2006), we
classify states of such systems into equivalence classes under orthonormal
basis changes. Next, we analyze the topological properties of the
corresponding MEN, and what structural changes they undergo once we
instantiate (measure) a component qubit. \newline (Duer, Vidal, and Cirac
2006) classifies three-qubit system states into fully entangled
(\textit{i.e.,} states of composite systems, in which no component system is
separable) and not fully entangled. Each of these classes is comprised of two
subclasses. The first one contains GHZ-like and $W$-like states, while the
second contains completely separable (\textit{i.e.,} there is no entanglement
present) and bi-separable states (\textit{i.e.,} two qubits are entangled and
separable from the remaining one). The classification is robust with respect
to local operations: in particular, by means of basis changes no state in one
class can be obtained from a state in a different class. It is a natural
question whether these four classes of three-qubit states can be characterized
purely in terms of their MEN topology.\vspace{1mm}\newline All possible MEN topologies corresponding to the four
classes of three-qubit states are summarized in Figure 3. One can easily see
that the MEN topology of all classes except for GHZ-like is invariant under basis
changes. However, in the case of GHZ-like states, there exist bases in which the
MEN graph assumes each of the configurations in Figure 3(d) (\textit{i.e.} the
complete graph, or any permutation of a three-node chain). In passing, we
observe that conditional separabilities, unlike unconditional ones, are not
preserved under basis changes. We conclude that the four classes of
three-qubit system states introduced above are completely characterized by the
topology of their associated MEN. \vspace{1mm}\newline Next, we consider the
issue of measurement and its impact on MEN topology. Measuring (instantiating)
a qubit collapses the wave function along one of the qubit's basis states,
destroying any entanglement with the remaining qubits. This has an impact on
MEN topology: specifically, after measuring a qubit, the MEN graph in the new
state is obtained by erasing all the edges which involve the measured qubit.
While measurement may destroy edges, it never introduces new ones, as
conditional separabilities are always preserved under qubit instantiation.

\begin{center}
\begin{figure}[ptb]
\begin{center}
\includegraphics[
					width=2.9603in]
					{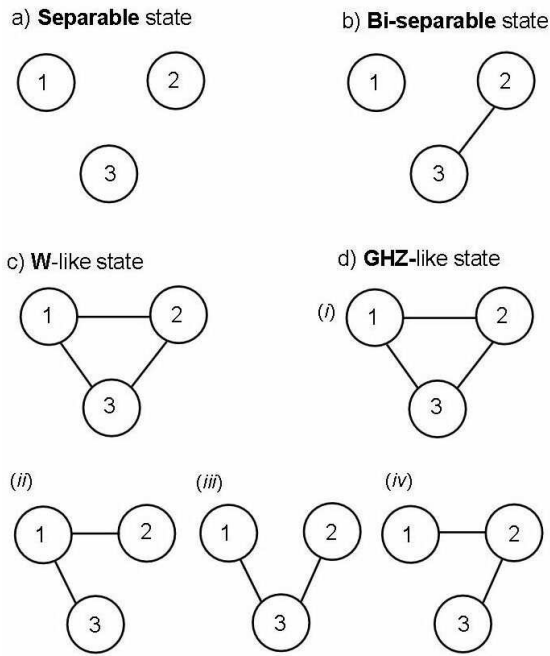}
\end{center}
\caption{MEN for all classes of three-qubit states.}%
\end{figure}
\end{center}

\section{Conclusions}

We introduced Markovian Entanglement Networks, a novel class of graphical
representations for quantum-mechanical systems, and argued that the modular
representation of quantum-mechanical states gives rise to computational
advantages analogous to those afforded by classical Markovian Networks. We
also showed that, based on MEN topology alone, one can effectively classify
all possible three-qubit system states. The most related approaches in the
literature are (Tucci 1997) and (Buzek and Plesch 2003). (Tucci 1997)
introduced \textit{Quantum Bayesian Networks (QBN)}, a graphical
representation based on directed acyclic graphs. In QBN the emphasis is on
modeling, rather than on computation, and QBN cannot be directly used to
facilitate computational tasks. (Buzek and Plesch 2003) introduced
\textit{Entangled Graphs (EG),} which capture the distribution of bipartite
entanglement in multi-qubit systems. The graphical decomposition in EG is based
on concurrence (a statistical measure of bipartite entanglement), and cannot
directly represent conditional separabilities. Furthermore, missing a modular
representation of amplitudes, EG cannot encode all the information contained
in a quantum-mechanical state.\newline Two main areas of application for MEN
come to mind: the classical simulation of quantum-mechanical systems, and
quantum computing. In particular, we hope that the identification and
exploitation of conditional separabilities in quantum-mechanical systems may
lead to the design of more efficient classical simulations in a variety of
computationally demanding applications, from physical chemistry to quantum
optics. Furthermore, in the context of quantum computation, we hope that the
graphical structure of MEN\ may provide a guide for the efficient design and
complexity analysis of quantum algorithms and circuits.

\section{References}

Duer, W., Vidal, G., Cirac, J. I., 2006. \textit{Three qubits can be entangled
in two inequivalent ways,} quant-ph/0005115. \newline \newline La Mura, P.,
Shoham, Y. 1999, \textit{Expected Utility Networks.} In \textit{Proceedings of
the 15th Conference on Uncertainty in Artificial Intelligence}, Stokholm.
\newline \newline Nielsen, M. A., and Chuang, I. L. eds. 2000, \textit{Quantum
Computation and Quantum Information.} Cambridge University Press.
\newline \newline Pearl, J., eds. 1988, \textit{Probabilistic Reasoning in
Intelligent Systems.} Morgan Kaufmann. \newline \newline Pearl, J., Paz, A.,
eds. 1987, \textit{Graphoids: a Graph-Based Logic for Reasoning about
Relevance Relations.} Advance in Artificial Intelligence-II, North-Holland
Publishing Company. \newline \newline Plesch M., Buzek V., 2003,
\textit{Entangled graphs: Bipartite entanglement in multi-qubit systems.}
Physical Review A \textbf{67}, 012322 \newline \newline Tucci, R., R., 1997,
\textit{Quantum Bayesian Nets}, quant-ph/9706039.
\end{document}